\documentclass[modern]{aastex62}

\usepackage{braket}
\usepackage{amsmath}
\usepackage{bm}

\newcommand{\acronym}[1]{{\small{#1}}}
\newcommand{\project}[1]{\textsl{#1}}
\newcommand{\apogee}{\project{\acronym{APOGEE}}}
\newcommand{\gaia}{\project{Gaia}}
\newcommand{\wise}{\project{\acronym{WISE}}}
\newcommand{\zmass}{\project{\acronym{2MASS}}}

\defcitealias{HER2018}{Paper I}

\revised{\today}
\submitjournal{ApJ}

\shorttitle{The Circular Velocity Curve of the Milky Way}
\shortauthors{Eilers et al.}

\begin{document}

\title{\textbf{The Circular Velocity Curve of the Milky Way from $5$ to $25$~kpc}}

\author[0000-0003-2895-6218]{Anna-Christina Eilers}
\affiliation{Max-Planck-Institute for Astronomy, K\"onigstuhl 17, 69117 Heidelberg, Germany}
\affiliation{International Max Planck Research School for Astronomy \& Cosmic Physics at the University of Heidelberg}

\author[0000-0003-2866-9403]{David W. Hogg}
\affiliation{Max-Planck-Institute for Astronomy, K\"onigstuhl 17, 69117 Heidelberg, Germany}
\affiliation{Center for Cosmology and Particle Physics, Department of Physics, New York University, 726 Broadway, New York, NY 10003, USA}
\affiliation{Center for Data Science, New York University, 60 Fifth Ave, New York, NY 10011}
\affiliation{Center for Computational Astrophysics, Flatiron Institute, 162 Fifth Ave, New York, NY 10010, USA}

\author[0000-0003-4996-9069]{Hans-Walter Rix}
\affiliation{Max-Planck-Institute for Astronomy, K\"onigstuhl 17, 69117 Heidelberg, Germany}

\author{Melissa K. Ness}
\affiliation{Center for Computational Astrophysics, Flatiron Institute, 162 Fifth Ave, New York, NY 10010, USA}
\affiliation{Department of Astronomy, Columbia University, 550W
120th St, New York, NY 10027, USA}


\correspondingauthor{Anna-Christina Eilers}
\email{eilers@mpia.de}

\begin{abstract}
We measure the circular velocity curve $v_{\rm c}(R)$ of the Milky Way with the highest precision to date across Galactocentric distances of $5\leq R \leq 25$~kpc. Our analysis draws on the $6$-dimensional phase-space coordinates of $\gtrsim 23,000$ luminous red-giant stars, for which we previously determined precise parallaxes using a data-driven model that combines spectral data from \apogee\ with photometric information from \wise, \zmass, and \gaia. We derive the circular velocity curve with the Jeans equation assuming an axisymmetric gravitational potential. At the location of the Sun we determine the circular velocity with its formal uncertainty to be $v_{\rm c}(R_{\odot}) = (229.0\pm0.2)\rm\,km\,s^{-1}$ with systematic uncertainties at the $\sim 2-5\%$ level. We find that the velocity curve is gently but significantly declining at $(-1.7\pm0.1)\rm\,km\,s^{-1}\,kpc^{-1}$, with a systematic uncertainty of $0.46\rm\,km\,s^{-1}\,kpc^{-1}$, beyond the inner $5$~kpc. We exclude the inner $5$ kpc from our analysis due to the presence of the Galactic bar, which strongly influences the kinematic structure and requires modeling in a non-axisymmetric potential. Combining our results with external measurements of the mass distribution for the baryonic components of the Milky Way from other studies, we estimate the Galaxy's dark halo mass within the virial radius to be $M_{\rm vir} = (7.25\pm0.26)\cdot 10^{11}M_{\odot}$ and a local dark matter density of $\rho_{\rm dm}(R_{\odot}) = 0.30\pm0.03\,\rm GeV\,cm^{-3}$. 
\end{abstract}


\section{Introduction}

The circular velocity curve of the Milky Way $v_{\rm c}(R)$ is the Galactocentric rest-frame velocity with which test particles would move on circular orbits at radius $R$ from the Galactic center, in an axisymmetric gravitational potential $\Phi$ of a disk galaxy, such as the Milky Way, i.e.
\begin{equation}
v_{\rm c}^2(R) = R\frac{\partial\Phi}{\partial R} \bigg\rvert_{z\approx 0}, 
\end{equation}
where $z$ denotes the height above the Galactic plane. 
This $v_{\rm c}(R)$ and in particular its value at the Sun's Galactocentric radius $R_{\odot}$, provide important constraints on the mass distribution of our Galaxy and the local dark matter density. The latter is crucial for interpreting and analyzing any direct as well as indirect detection experiments of dark matter, whereas the shape of the rotation curve is a fundamental parameter for models of the Galactic disk. The local circular velocity at the Sun's location plays an important role when placing the Milky Way in a cosmological context and asking for instance, whether it falls onto the Tully-Fisher relation \citep{Klypin2002}. 


Previously, the Sun's circular motion has been inferred by measuring its velocity with respect to an object at rest within our Galaxy, such as Sgr A* \citep{ReidBrunthaler2004}. However, this method only determines the tangential velocity of the Sun, which differs from the circular velocity of the Galaxy at the location of the Sun by the Sun's so-called peculiar velocity. The method also requires knowledge of the Sun's Galactocentric distance, the precision of which improved significantly by the recent result from the \citet{gravity2018} \citep[see][for a review]{BlandHawthornGerhard2016}. 
An alternative way to determine the circular velocity at the Sun's radius, which does not require a precise estimate of the Sun's location with respect to the Galaxy, is the  modelling of tidal streams in the Galactic gravitational potential, which has been analyzed with the GD-1 stream \citep{Koposov2010} and Palomar 5 \citep{Kuepper2015}. 

For the inner region of the Galaxy the shape of the circular velocity curve has been inferred via the tangent-point method, based on the observed kinematics of the \ion{H}{1} or $\rm CO$ emission of the interstellar medium, under the assumption of purely circular orbits of the gas \citep{Gunn1979, Fich1989, Levine2008}.  
The circular velocity in the outer part of the Galaxy, i.e. outside of the Galactocentric radius of the Sun, has been measured by means of line-of-sight velocities and distances using a variety of tracers, such as the thickness of the \ion{H}{1} layer \citep{Merrifield1992},  spectrophotometric distances of \ion{H}{2} regions combined with radial velocities of associated molecular clouds \citep{Fich1989, BrandBlitz1993}, radial velocities measurements of planetary nebulae \citep{SchneiderTerzian1983}, classical cepheids \citep{Pont1997} or RR Lyrae stars \citep{Ablimit2017, Wegg2018}, blue horizontal branch stars in the halo \citep{Xue2009, Kafle2012}, red giant branch and red clump stars stars in the Galactic disk \citep{Bovy2012, Huang2016}, or masers in high mass star forming regions \citep{Reid2014}. 
However, such modeling requires a numerous and luminous tracer population and has significant limitations if the chosen tracers are either quite rare, e.g. classical cepheids, or not very luminous and hence not observable to large distances, e.g. red clump stars. Additionally, uncertainties in the distance estimates can lead to significant biases in the analysis of the circular velocity curve.

In this paper we present a new measurement of the Milky Way's circular velocity derived from the $6$-dimensional phase-space measurements for $\gtrsim 23,000$ red-giant stars at Galactocentric distances between $5\lesssim R\lesssim 25$~kpc. Critical for such a measurement are precise and accurate distances to luminous tracers that can be observed over a wide range of Galactic distances. Stars at the top of the red-giant branch (RGB) are frequent and very bright and hence well suited as a tracer population for this study. In \citet[hereafter \citetalias{HER2018}]{HER2018} we developed a data-driven model making use of spectral data from \apogee\ DR14 as well as photometric information from \gaia, \zmass, and \wise\ to improve on the measured parallaxes from \gaia~DR2, deriving parallaxes to all stars with only $\sim10\%$ uncertainties. 

We then derive the Milky Way's circular velocity within the Galactic plane based on the Jeans equation, which relates the circular velocity to the number density and Galactocentric radial as well as tangential velocity dispersions of the tracer population, assuming an axisymmetric gravitational potential for the Milky Way \citep[see also e.g.][]{Bhattacharjee2014}. 

The outline of this paper is as follows: in \S~\ref{sec:data} we briefly present the data set of red giant stars and summarize the model with which we previously derived spectrophotometric distances to these stars in \S~\ref{sec:distance}. We introduce the Jeans equation in \S~\ref{sec:model} and present the resulting rotation curve in \S~\ref{sec:results}, before summarizing our results in \S~\ref{sec:summary}. 


\section{Data Set}\label{sec:data}


In order to determine $v_{\rm c}(R)$, we need to infer the Galactocentric radius $R$, the tangential velocity $v_{\varphi}(R)$, the radial velocity $v_{R}(R)$, and their uncertainties for each star from the $6$-dimensional phase-space information, which requires the knowledge of precise distances to these stars. 
In \citetalias{HER2018} we infer parallaxes with a purely linear model, which we briefly summarize in \S~\ref{sec:distance}, and hence we restrict ourselves to a very limited region of the stellar parameter space. We choose stars on the upper red giant branch with low surface-gravity, i.e. with a surface gravity of $0\leq \log g \leq 2.2$, which selects stars that are more luminous than the red clump. Furthermore, we select stars which have existing spectral data from \apogee\ DR14 \citep{Majewski2017}, as well as complete photometric information in the $G$-band, $G_{\rm BP}$ and $G_{\rm RP}$ from \gaia\ DR2 \citep{Gaia2016}, near-infrared data in the $J$-, $H$-, and $K$-band  from \zmass\ \citep{2MASS}, and $W1$ and $W2$ (at $3.6\,\mu$m and $4.5\,\mu$m, respectively) from \wise\ \citep{WISE}. After applying further data quality cuts \citepalias[see][for details]{HER2018}, this results in $44,784$ luminous red giant stars with complete spectral and photometric data, suitable for precise parallax and hence $6$-dimensional phase-space estimates. 

For the analysis of the circular velocity curve we only consider stars within a wedge of $60^{\circ}$ from the Galactic center towards the direction of the Sun, i.e. $\pm30^{\circ}$. Additionally, we apply a cut on the height above the Galactic plane in order to restrict ourselves to stars within the disk, i.e. we take all stars which satisfy either $|z| \leq 0.5$~kpc or lie within $6^{\circ}$ from the Galactic plane, i.e. $|z|/R \leq \tan(6^{\circ})$. In order to avoid contamination from halo stars we also exclude stars with a vertical velocity component of $|v_z|>100\,\rm km\,s^{-1}$. 
Furthermore, we select stars with low $\alpha$-element abundances, i.e. $[\alpha/\rm Fe] < 0.12$, in order to avoid large asymmetric drift corrections, which results in a total of $23,129$ giant stars for the analysis of $v_{\rm c}(R)$. \\

\section{Spectrophotometric Parallaxes}\label{sec:distance}

In \citetalias{HER2018} we developed a data-driven, linear model combining only spectroscopic data from \apogee\ DR14 and photometry from \gaia\ DR2, \zmass, and \wise\ to determine precise parallaxes to stars located at the upper end of the red giant branch, which we will summarize here briefly. 
Our method assumes that red giant stars are dust-correctable, standardizable candles, which means that we can infer their distance modulus -- and thus their parallax -- from their spectroscopic and photometric features. 


For each star $n$ we create a $D$-dimensional feature vector $x_n$ that contains the chosen stellar features. 
We then optimize the log-likelihood function, which can be expressed as
\begin{equation}
\ln\mathcal{L}(\theta) = -\frac{1}{2}\chi^2(\theta) \hspace{.5 cm}{\rm with} \hspace{.5 cm}\chi^2(\theta) \equiv \sum_{n=1}^N \frac{[\varpi^{\rm (a)}_n - \exp(\theta\cdot x_n)]^2}{\sigma_n^2}, \label{eq:likelihood}
\end{equation}
where $\varpi^{\rm (a)}_n$ and $\sigma_n$ are the astrometric parallax measurements and their uncertainties from \gaia, and $\theta$ is the $D$-dimensional vector that we optimize for, containing the linear coefficients of our model. 
%
Once the model is optimized for a training set, the output allows us to predict the so-called \textit{spectrophotometric parallax} $\varpi^{\rm (sp)}$ for all stars in the disjoint validation set. 
Our model predicts the spectrophotometric parallaxes with $\sim 10\%$ uncertainties, and the resulting distances are more than twice as precise as \gaia's predictions at heliocentric distances of $\gtrsim3$~kpc ($\gtrsim1$~kpc) for stars with $G\sim12\rm \, mag$ ($G\sim14\rm \, mag$). At $\sim 15$~kpc distance from the Sun, the derived spectrophotometric distances are a factor of $\approx 6-8$ times more informative and thus they enable us to make precise maps of the Milky Way \citepalias[see][]{HER2018}. 

Afterwards, we transform all stars into the Galactocentric coordinate frame making use of the barycentric radial velocities from \apogee, as well as the distance from the Sun to the Galactic center $R_{\odot} = 8.122 \pm 0.031$~kpc \citep{gravity2018}, its height above the Galactic plane $z_{\odot}\approx0.025$~kpc \citep{Juric2008}, and the Galactocentric velocity of the Sun, i.e. $v_{\odot,x}\approx -11.1\,\rm km\,s^{-1}$,  $v_{\odot,y}\approx 245.8\,\rm km\,s^{-1}$, and $v_{\odot,z}\approx 7.8\,\rm km\,s^{-1}$, derived from the proper motion of Sgr A*, which we assume to be the rest-frame of the Galactic center \citep{ReidBrunthaler2004}. Note that our method does not require any knowledge of the local standard of rest. 

\section{Jeans Modelling of the Circular Velocity}\label{sec:model}

Assuming an axisymmetric gravitational potential of the Milky Way, we solve the cylindrical form of the Jeans equation \citep{Jeans1915, BinneyTremaine}, which links the moments of the velocity distribution and the density of a collective of stars to the gravitational potential, i.e. 
\begin{equation}
\frac{\partial \nu \langle v^2_R\rangle}{\partial R} + \frac{\partial\nu\langle v_R v_z\rangle}{\partial z} + \nu\left(\frac{\langle v^2_R\rangle - \langle v^2_{\varphi}\rangle}{R} + \frac{\partial\Phi}{\partial R} \right) = 0,  \label{eq:jeans}
\end{equation}
where 
$\nu$ denotes the density distribution of the tracer population. We can then derive the circular velocity via
\begin{equation}
v^2_{\rm c}(R) = \langle v^2_{\varphi}\rangle - \langle v^2_R\rangle\left(1 + \frac{\partial \ln \nu}{\partial \ln R} + \frac{\partial \ln \langle v^2_R\rangle}{\partial \ln R}\right). \label{eq:vcirc2}
\end{equation}
Note that for deriving Eqn.~\ref{eq:vcirc2} we neglected the second term in Eqn.~\ref{eq:jeans}, because the cross-term $\langle v_R v_{\varphi}\rangle$ and its vertical gradient is $\sim 2-3$ orders of magnitude smaller compared to the remaining terms out to Galactocentric distances of $R\sim 18$~kpc and hence effects the circular velocity at the $\sim 1\%$ level (see \S~\ref{sec:sys}). 

The terms in Eqn.~\ref{eq:vcirc2} are estimated as follows:
In the presence of measurement uncertainties $C_v$ we calculate the velocity tensor $V$ by assigning
\begin{equation}
V \leftarrow \langle v v^T\rangle - C_v, 
\end{equation}
where $ \langle v v^T\rangle$ denotes the squared velocity averaged over an ensemble of stars, and replace the respective terms in Eqn.~\ref{eq:vcirc2} with:
\begin{align*}
\langle v^2_{R}\rangle &\leftarrow V_{RR},\\
\langle v^2_{\varphi}\rangle &\leftarrow V_{\varphi\varphi}. 
\end{align*}

To use Eqn.~\ref{eq:vcirc2} we need to know the radial density profile for the tracer population. Because we do not know the selection function of \apogee\ precisely enough to determine the profile from the data itself, we simply and plausibly chose an exponential function, i.e. 
\begin{equation}
\nu(R)\propto \exp\left(-\frac{R}{R_{\rm exp}}\right)\label{eq:tracer_density}
\end{equation}
with a scale length of $R_{\rm exp} = 3$~kpc, which is in good agreement with previous studies \citep{BlandHawthornGerhard2016}. 
However, the chosen value of the scale length is the dominant source of systematic uncertainty on the circular velocity curve and causes errors at the $\sim2\%$ level, which we discuss in detail in \S~\ref{sec:sys}. 

\begin{figure*}[t!]
\centering
\includegraphics[width=\textwidth]{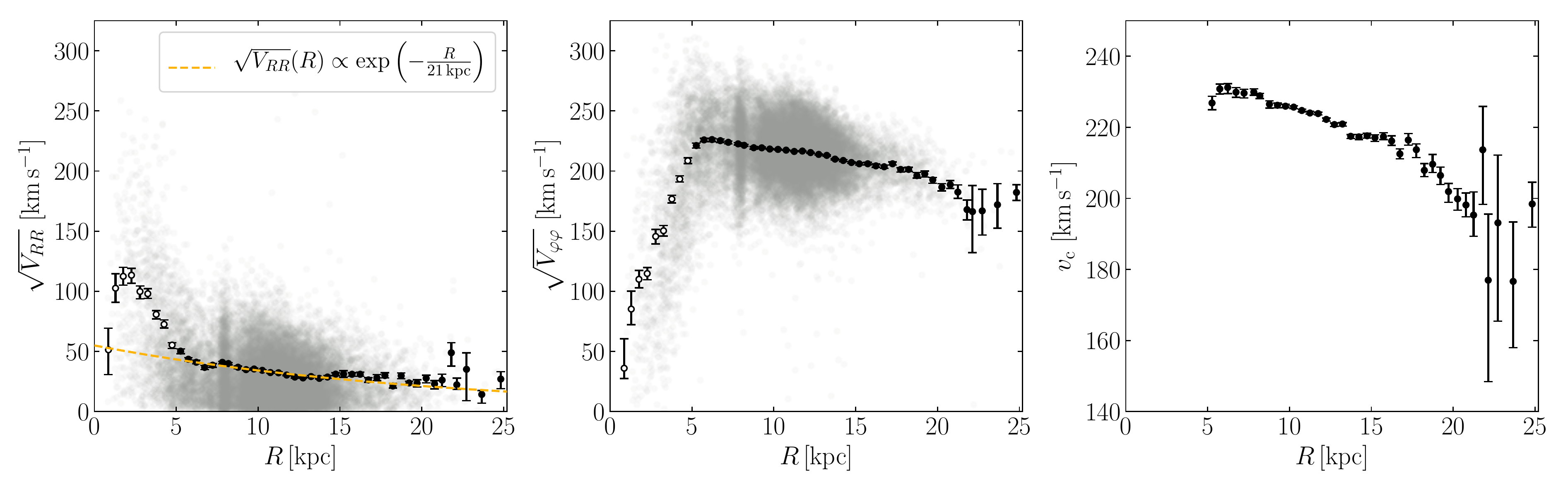}
\caption{Radial profiles of the components of the velocity tensor $\sqrt{V_{RR}}$ (left panel), $\sqrt{V_{\varphi\varphi}}$ (middle panel), and the circular velocity $v_{\rm c}$ (right panel). The small grey dots show individual stars, whereas the black data points show the ensemble averaged values with uncertainties determined via bootstrapping with $100$ samples. We exclude all stars within a distance of $R < 5$~kpc from the Galactic center (open data points) from our analysis due to the presence of the Galactic bar. 
\label{fig:radial}} 
\end{figure*}

In order to estimate of the radial dependence of the velocity tensor and to infer the Milky Way's circular velocity, we calculate the radial profiles of the velocity tensor components by averaging all stars within annuli with a minimum width of $\Delta R = 0.5$~kpc or a minimum number of three stars per bin (Fig.~\ref{fig:radial}). 
We then obtain the estimate $\sqrt{V_{RR}}(R)$ from the data itself. We subsequently model this dependency with an exponential function and find a scale length of $R'_{\rm exp} \approx 21$~kpc beyond the inner $5$~kpc (yellow dashed curve in the left panel of Fig.~\ref{fig:radial}). 

Note that within a distance of $R \lesssim 5$~kpc we observe more complex, non-axisymmetric dynamics due to the presence of the bar in the Milky Way. The modeling of the orbits of stars within this region will require to account for deviations from an axisymmetric gravitational potential. However, this is beyond the scope of this paper and hence we exclude this region from our analysis. 
We can now calculate $v_{\rm c}(R)$ via Eqn.~\ref{eq:vcirc2}, which is shown in the right panel of Fig.~\ref{fig:radial}. 

\begin{figure*}[t!]
\centering
\includegraphics[width=\textwidth]{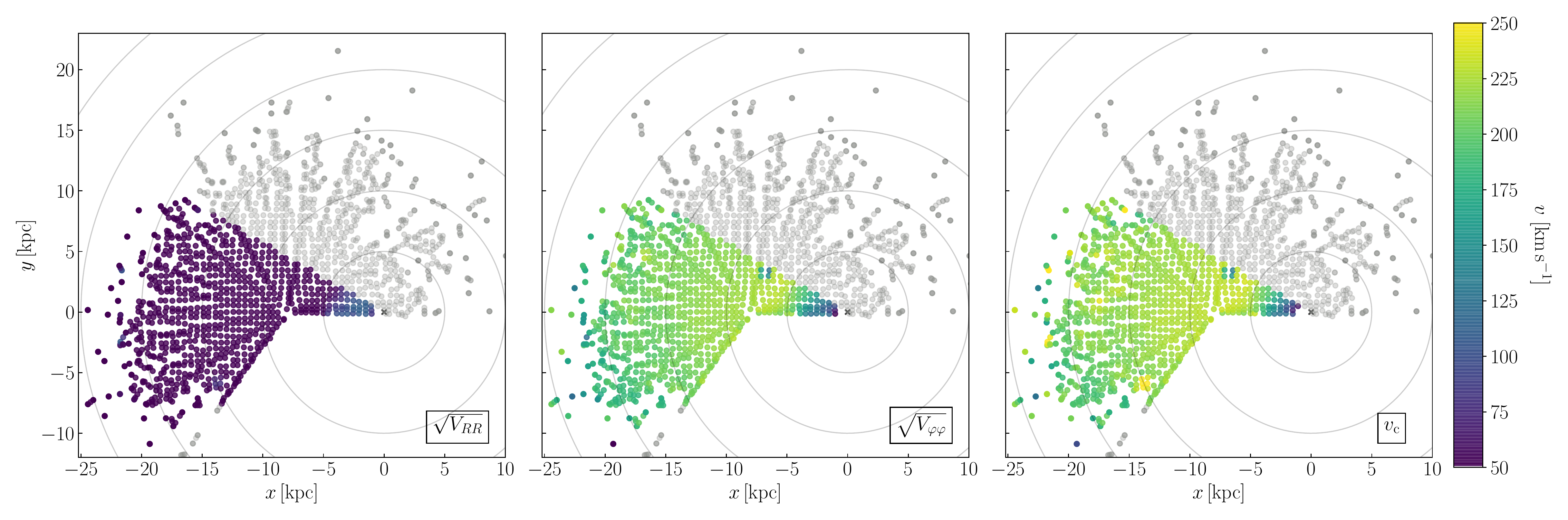}
\caption{Maps of the Milky Way colored by the components of the velocity tensor $\sqrt{V_{RR}}$ (left panel), $\sqrt{V_{\varphi\varphi}}$ (middle panel), and $v_{\rm c}$ (right panel). Each dot represents an average of the ensemble of stars located within $1$~kpc$^2$ in the $x-$ and $y-$direction. Stars in the grey region are not taken into account for our analysis of the circular velocity curve. \label{fig:maps}} 
\end{figure*}

The maps of the Milky Way colored by the different components of the velocity tensor, as well as the resulting circular velocity are shown in Fig.~\ref{fig:maps}.

\section{Results}\label{sec:results}

\subsection{The Circular Velocity Curve}\label{sec:v_circ}

The resulting curve of the Milky Way's circular velocity $v_c(R)$ is shown in Fig.~\ref{fig:vrot}, while the individual measurements are listed in Table \ref{tab:measurements}. It covers a large radial extent between $5\lesssim R\lesssim 25$~kpc and is very precisely measured, with average uncertainties of $\sigma_{v_c}\lesssim3\,\rm km\,s^{-1}$ determined via bootstrapping from $100$ samples. The circular velocity curve shows a gentle but significant decline with increasing radius and can be well approximated by a linear function:
\begin{equation}
v_{\rm c}(R) = (229.0 \pm 0.2) {\rm\,km\,s^{-1}} - (1.7\pm0.1) {\rm\,km\,s^{-1}\,kpc^{-1}} \cdot(R - R_{\odot}),\label{eq:linear_fit} 
\end{equation}
which implies a circular velocity with a formal uncertainty of the fit of $v_{\rm c}(R_{\odot}) = (229.0 \pm 0.2) {\rm\,km\,s^{-1}}$ at the position of the Sun with a derivative of $(-1.7\pm0.1)\rm\,km\,s^{-1}\,kpc^{-1}$, indicating a constantly, gently declining circular velocity curve. 

\subsection{Systematic Uncertainties}\label{sec:sys}

\begin{figure}
\centering
\includegraphics[width=\textwidth]{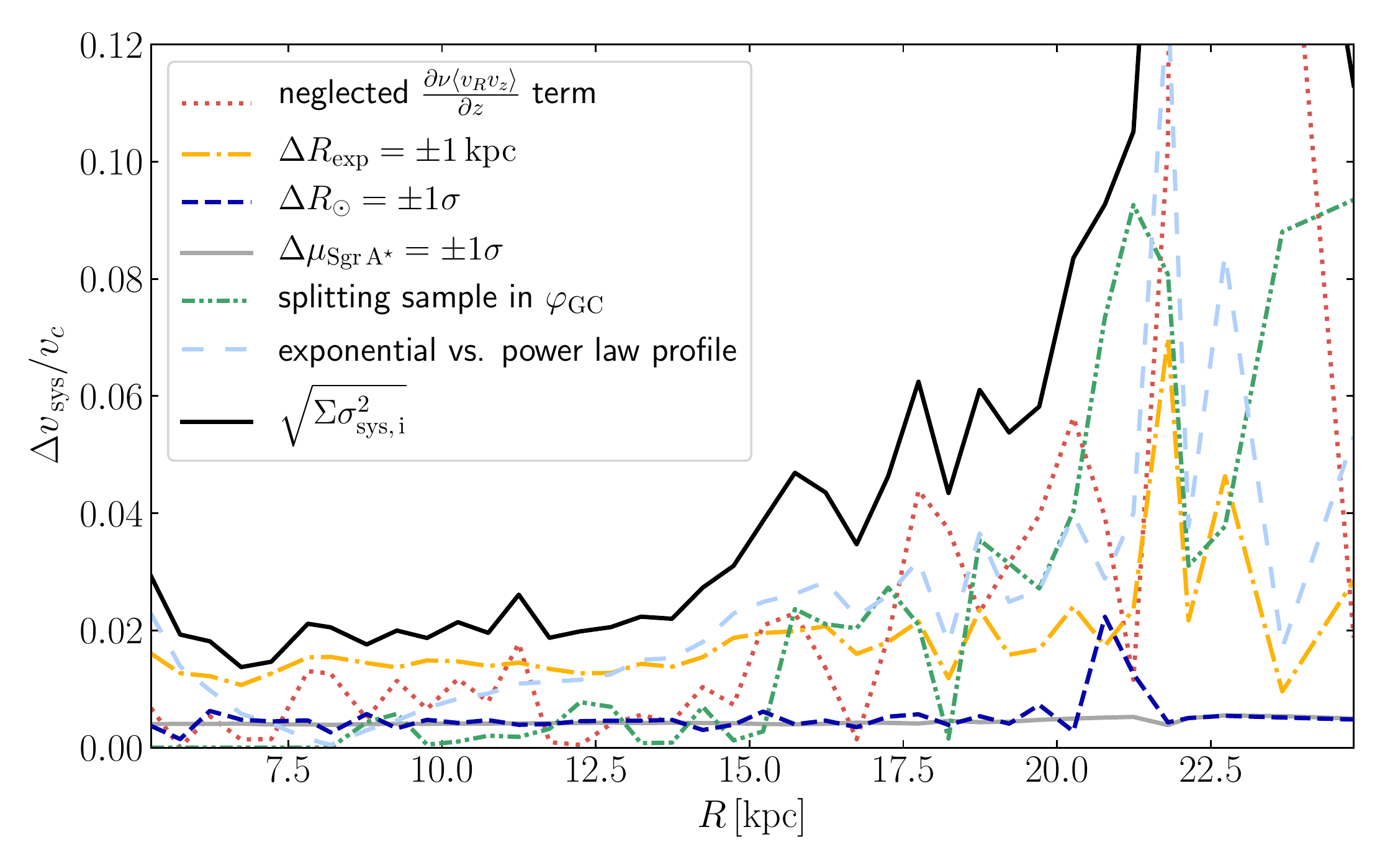}
\caption{Summary of potential systematic uncertainties in the circular velocity curve. We plot the relative deviation in the circular velocity $\Delta v_{\rm sys}$ to our fiducial circular velocity curve $v_c$, dependent on the Galactocentric radius. We estimate systematic uncertainties arising from the neglected term in Eqn.~(\ref{eq:jeans}) (red dotted curve), from varying the exponential scale length of the density profile of the tracer population (yellow dash-dotted curve), and from changing this density profile from an exponential function to a power law (light blue dashed curve) with an index $\alpha = -2.7$, which has the same slope at the location of the Sun as our fiducial exponential model. Very moderate systematic uncertainties at the $\lesssim 1\%$ level arise from splitting the data sample into two distinct wedges (green dash-dotted curve, see \S~\ref{sec:sys}), varying the distance from the Sun to the Galactic center (blue dashed curve), 
or uncertainties in the proper motion of Sgr $\rm A^{*}$ (grey curve). 
All systematic uncertainties added up (black curve) result in uncertainties in $v_c$ at the $\sim 2-5\%$ level out to $R\sim 20\,\rm kpc$.  
\label{fig:sys}}
\end{figure}

Various systematic uncertainties influence our measurement of the circular velocity curve. In Fig.~\ref{fig:sys} we illustrate the relative systematic differences in the circular velocity $\Delta v_{\rm sys}/v_{\rm c}$ as a function of the Galactocentric radius. 

As already mentioned, the largest contribution to the systematic uncertainty of the circular velocity at the location of Sun stems from the unknown density profile of the tracer population. Our fiducial model assumes an exponential function for this density profile with an exponential scale length of $R_{\rm exp}=3\,\rm kpc$. Varying this scale length, i.e. $\Delta R_{\rm exp} = \pm 1\,\rm kpc$, causes systematic uncertainties at the $\sim 2\%$ level. We also tested the systematic uncertainties arising from the chosen functional form for this density profile. If we instead apply a density profile following a power law, for which we chose an index of $\alpha=-2.7$, which has the same slope as our fiducial exponential function at the location of the Sun, we obtain a systematic error that increases with distance from the Sun to $2-3\%$ for $R\lesssim 20\,\rm kpc$. 

For calculating the circular velocity curve we neglected the term $\frac{\partial\nu\langle v_Rv_z\rangle}{\partial z}$ in the Jeans equation (Eqn.~\ref{eq:jeans}), which also adds a systematic uncertainty. We estimated this uncertainty to be at the $1\%$ level out to $R\sim 18\,\rm kpc$, but this term might have larger systematic effects at larger Galactocentric radii. 

In order to estimate additional systematic uncertainties on the circular velocity curve arising from our data sample, we split the region, i.e. the $60^{\circ}$ wedge, within which we consider stars for the analysis, into two disjoint smaller wedges of $30^{\circ}$ each, and perform the same analysis with both sets of stars. We estimate the systematic uncertainties on the circular velocity by the difference between the resulting fit parameters from the two disjoint data sets. For the circular velocity at the location of the sun we determine a systematic error of $\sigma_{v_c(R_{\odot})} \approx 2.02\, \rm km\,s^{-1}$, which corresponds to uncertainties at the $\lesssim 1\%$ level, whereas the systematic error on the slope of the rotation curve is approximately $0.46 \rm\,km\,s^{-1}\,kpc^{-1}$, which adds uncertainties at the $\sim 27\%$ level. 

Further minor systematic uncertainties on the circular velocity at the $\lesssim 1\%$ level stem from the uncertainty on the Galactocentric distance of the Sun, i.e. $\sigma_{R_{\odot}}=31\rm\,pc$ \citep{gravity2018}, and the uncertainties in the proper motion of Sgr $\rm A^{\star}$, i.e. $\sigma_{\mu_l}=0.026\,\rm mas\,yr^{-1}$ and $\sigma_{\mu_b}=0.019\,\rm mas\,yr^{-1}$ \citep{ReidBrunthaler2004}. 

\begin{figure}
\centering
\includegraphics[width=\textwidth]{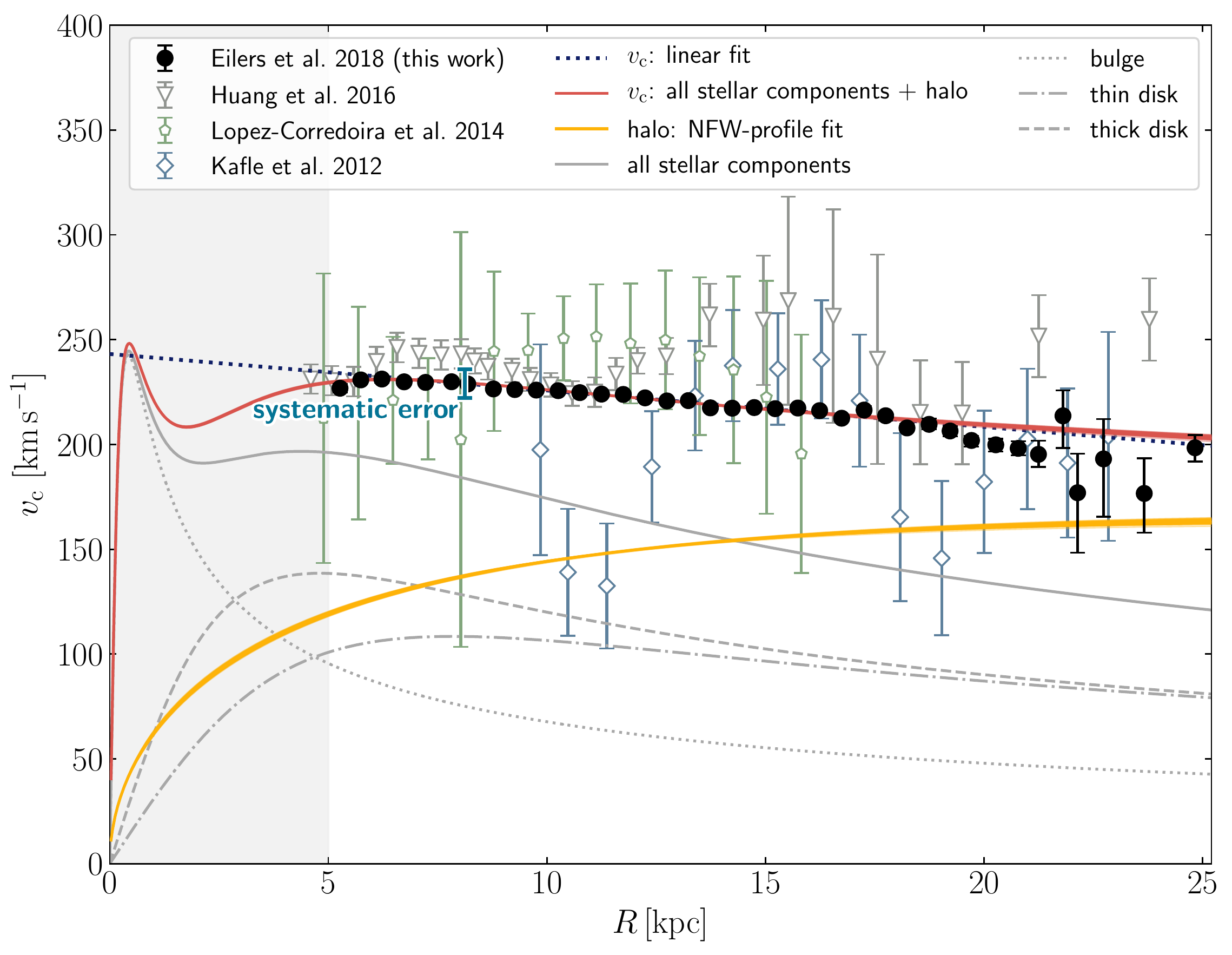}
\caption{The new measurements of the circular velocity curve of the Milky Way are shown as the black data points. The errorbars are estimated via bootstrapping and do not include any systematic uncertainties. We note the systematic error at the location of the Sun, which influences our results at the $\lesssim 3\%$ level (see \S~\ref{sec:sys}). The blue dotted curve shows a linear fit to our data (Eqn.~\ref{eq:linear_fit}), whereas the red curves show $100$ random draws from the posterior distribution of the fit parameters to the circular velocity modeled as a sum of stellar components, i.e. bulge, thin and thick disk (grey curves), and a dark matter halo estimated by an NFW-profile (yellow curves, also showing $100$ random draws from the posterior). The measurements of various other studies of the circular velocity are shown as colored data points. The light grey shaded region marks the region, where dynamics are strongly influenced by the Milky Way's bar. \label{fig:vrot}}
\end{figure}

All systematic uncertainties taken together affect our measurement of the circular velocity at the $2-5\%$ level out to $R\sim 20\,\rm kpc$. At Galactocentric distances beyond $20\,\rm kpc$ the systematic error rises, since our data set only contains relatively few stars at these distances and is thus not very constraining. At the location of the Sun we estimate a systematic uncertainty of $\lesssim 3\%$, which is marked in Fig.~\ref{fig:vrot}. 

We would like to point out that the systematic uncertainties cannot eliminate the gentle decline in the circular velocity curve that we find. In order to obtain a flat circular velocity curve, the exponential scale length of the density profile of the tracer population would has to be very small, i.e. $R_{\rm exp} < 1$~kpc. Such small scale lengths, however, would imply that almost no stars would be located at Galactocentric distances of $R\sim 20$~kpc and beyond, which is clearly not the case. 

\startlongtable
\begin{deluxetable*}{RRRR}
\tabletypesize{\footnotesize}
\tablecaption{Measurements of the circular velocity of the Milky Way.  \label{tab:measurements}}
\tablehead{\dcolhead{R\,\rm [kpc]} & \dcolhead{v_{\rm c}\,\rm [km\,s^{-1}]} & \dcolhead{\sigma^{-}_{v_{\rm c}}\,\rm [km\,s^{-1}]} & \dcolhead{\sigma^{+}_{v_{\rm c}}\,\rm [km\,s^{-1}]}}
\startdata
5.27 & 226.83 & 1.91 & 1.90 \\
5.74 & 230.80 & 1.43 & 1.35 \\
6.23 & 231.20 & 1.70 & 1.10 \\
6.73 & 229.88 & 1.44 & 1.32 \\
7.22 & 229.61 & 1.37 & 1.11 \\
7.82 & 229.91 & 0.92 & 0.88 \\
8.19 & 228.86 & 0.80 & 0.67 \\
8.78 & 226.50 & 1.07 & 0.95 \\
9.27 & 226.20 & 0.72 & 0.62 \\
9.76 & 225.94 & 0.42 & 0.52 \\
10.26 & 225.68 & 0.44 & 0.40 \\
10.75 & 224.73 & 0.38 & 0.41 \\
11.25 & 224.02 & 0.33 & 0.54 \\
11.75 & 223.86 & 0.40 & 0.39 \\
12.25 & 222.23 & 0.51 & 0.37 \\
12.74 & 220.77 & 0.54 & 0.46 \\
13.23 & 220.92 & 0.57 & 0.40 \\
13.74 & 217.47 & 0.64 & 0.51 \\
14.24 & 217.31 & 0.77 & 0.66 \\
14.74 & 217.60 & 0.65 & 0.68 \\
15.22 & 217.07 & 1.06 & 0.80 \\
15.74 & 217.38 & 0.84 & 1.07 \\
16.24 & 216.14 & 1.20 & 1.48 \\
16.74 & 212.52 & 1.39 & 1.43 \\
17.25 & 216.41 & 1.44 & 1.85 \\
17.75 & 213.70 & 2.22 & 1.65 \\
18.24 & 207.89 & 1.76 & 1.88 \\
18.74 & 209.60 & 2.31 & 2.77 \\
19.22 & 206.45 & 2.54 & 2.36 \\
19.71 & 201.91 & 2.99 & 2.26 \\
20.27 & 199.84 & 3.15 & 2.89 \\
20.78 & 198.14 & 3.33 & 3.37 \\
21.24 & 195.30 & 5.99 & 6.50 \\
21.80 & 213.67 & 15.38 & 12.18 \\
22.14 & 176.97 & 28.58 & 18.57 \\
22.73 & 193.11 & 27.64 & 19.05 \\
23.66 & 176.63 & 18.67 & 16.74 \\
24.82 & 198.42 & 6.50 & 6.12 \\
\enddata
\tablecomments{Columns show the Galactocentric radius, the circular velocity, and its negative and positive errorbars. }
\end{deluxetable*}

\subsection{Velocity Contribution from the Dark Matter Halo}

We now proceed to explore in a simple manner how the inferred $v_c(R)$ reflects contributions from different mass components. 
Due to the linearity of Poisson's equation we can approximate the Milky Way's gravitational potential as a sum of potentials evoked by its individual components $i$, i.e. the bulge, thin and thick disk, and the halo, hence
\begin{align}
\Phi = \sum_{i}\Phi_i \Rightarrow v^2_{\rm c} = \sum_{i} v^2_{\rm c, i}. 
\end{align}
In this initial analysis, we assume that the stellar components are very well determined, and fit only the velocity contribution of the dark matter halo, for which we assume a gravitational potential approximated by a Navarro-Frenk-White profile \citep[NFW;][]{Navarro1997}. For the gravitational potentials of the thin and thick disk we assume Miyamoto-Nagai profiles \citep{MiyamotoNagai1975}, and for the bulge we assume a spherical Plummer potential \citep{Plummer1911}, while adapting the parameter values of the enclosed mass, the scale length, and the scale height from \citet[][model I]{Pouliasis2017}. 

We apply the Markov Chain Monte Carlo (MCMC) affine invariant sampler \texttt{emcee} \citep{emcee}, to fit the velocity contribution from the dark matter halo, assuming flat priors for the virial mass $M_{\rm vir}\in[10^{10}, 10^{15}]\,M_{\odot}$ and the concentration parameter $c\in[0, 50]$ of the NFW-profile. We adopt the mean of the posterior probability distribution as the best estimate and obtain $M_{\rm vir}=(7.25\pm0.25)\cdot10^{11}\,M_{\odot}$ and $c=12.8\pm0.3$, corresponding to a virial radius of $R_{\rm vir} = 189.3\pm2.2$~kpc, 
a scale radius of $R_s = 14.8\pm0.4$~kpc, and a characteristic density of $\rho_0 = (1.06\pm0.09) \cdot 10^7\,M_{\odot}\rm\,kpc^{-3}$. At the location of the Sun the implied dark matter density is $\rho_{\rm dm}(R_{\odot}) = 0.30\pm0.03\,\rm GeV\,cm^{-3}$, presuming the halo is spherical. 
Our estimate of the velocity contribution of the dark matter halo also reveals that the dynamics of the Milky Way are becoming dominated by dark matter beyond $R\gtrsim 14$~kpc, i.e. the largest contribution to the circular velocity curve originates from the dark matter halo, whereas the inner part is dominated by the stellar components. 


\section{Summary and Discussion}\label{sec:summary}

In this paper we presented new measurements of the Milky Way's circular velocity curve between $5\leq R\leq 25$~kpc. \gaia's precise $6$-dimensional phase space information for a large sample of tracer stars, together with the new measurements of the Galactocentric rest-frame \citep{gravity2018}, have enabled the most precise measurements of the circular velocity to date. Combining our equatorial circular velocity curve measurement with  the kinematics of the Galactic halo from stellar streams will enable new constraints of the $3$-dimensional profile and flattening of the Milky Way's halo \citep[e.g.][]{Xue2015}. 

We derive the circular velocity based on a Jeans model under the assumption of an axisymmetric gravitational potential. We used $\gtrsim 23,000$ stars at the upper red giant branch as a tracer population, which are well suited because they are frequent and very luminous, and hence observable over large Galactic distances. We select all stars close to the Galactic plane, because the Milky Way's disk is a cold and precise tracer of kinematic structure. 
We make use of the fact that red giant stars -- similar to red clump stars -- are standardizable candles, which allowed us to obtain precise spectrophotometric distance estimates from a linear data-driven model \citepalias{HER2018}. 

The dynamics of the inner $5$~kpc of the Milky Way disk are strongly influenced by the presence of the Milky Way bar and thus need to be modeled in a non-axisymmetric potential which is beyond the scope of this paper. Beyond $R\gtrsim5$~kpc, Jeans modelling yields a precise and robust estimate of the circular velocity curve, which shows a gently declining slope of $(-1.7\pm0.1)\rm\,km\,s^{-1}\,kpc^{-1}$, with a systematic uncertainty of $0.46\rm\,km\,s^{-1}\,kpc^{-1}$. Our result is in reasonably good agreement with another recent analysis of the Milky Way's circular velocity curve based on classical cepheids by \citet{Mroz2018}, who find a slope of $(-1.41\pm0.21)\rm\,km\,s^{-1}\,kpc^{-1}$. However, our derivative is significantly less shallow than previous studies by \citet{Bovy2012}, \citet{BovyRix2013} or \citet{Reid2014} suggest, who estimate a slope that is consistent with a flat circular velocity curve, which is excluded by our estimate with $>3\sigma$ significance. 

A declining circular velocity curve has not been observed in many other disk galaxies in the local universe, which rather show a flat or even increasing circular velocity curve \citep[e.g.][]{Rubin1980, Sofue1999}. Galaxies with declining circular velocity curves have yet only been reported at higher redshift. For instance, \citet{Genzel2017} studied six massive star-forming galaxies at $z\approx 2$ and found declining circular velocities curves, claiming that these galaxies are baryon-dominated and their dark matter content smaller than in disk galaxies in the local universe \citep[see also][]{Lang2017}. They argue that the observations suggest that baryons in the early universe during the peak epoch of star formation efficiently condense at the centers of dark matter halos when gas fractions are higher and dark matter is less concentrated. 


Our estimate of the circular velocity at the Sun's position of $v_{\rm c}(R_{\odot}) = (229.0\pm0.2)\rm\,km\,s^{-1}$ is very precisely determined, with systematic uncertainties at the $\sim 2-3\%$ level. 
Despite the formally very small uncertainties in our circular velocity curve, the systematic uncertainties introduced by various factors, such as the choice of the exponential scale length of the tracer population for instance (Eqn.~\ref{eq:tracer_density}), the assumed axisymmetry of the gravitational potential, or the \apogee\ selection function could bias our results. These systematic uncertainties might increase at large Galactocentric distances, i.e. $R\gtrsim 20$~kpc, where our data is not very constraining. 
Our estimate is in good agreement with the estimate by \citet{Koposov2010}, i.e. $v_{\rm c}(R_{\odot}) = 221^{+16}_{-20}\rm\,km\,s^{-1}$, and \citet{Wegg2018}, who estimated $v_{\rm c}(R_{\odot}) = (222\pm 6)\rm\,km\,s^{-1}$ from modeling RR Lyrae stars with \gaia. However, it is significantly lower than the measurement by \citet{Reid2014}, i.e. $v_{\rm c}(R_{\odot}) = (240\pm8)\rm\,km\,s^{-1}$, but higher than the estimate of $v_{\rm c}(R_{\odot})=(218\pm6)\rm\,km\,s^{-1}$ by \citet{Bovy2012}. We also calculated the circular velocity from combining the result for the Sun's Galactocentric distance by \citet{gravity2018} with the recently published Oort's constants by \citet{Bovy2017}, which he derived from main-sequence stars from the \gaia\ DR1 Tycho-Gaia Astrometric Solution (TGAS), and obtain $v_{\rm c}(R_{\odot})=(220.9\pm4.7)\rm\,km\,s^{-1}$, which is about $2\sigma$ lower than our estimate. 


The very smooth shape of our circular velocity curve is in broad agreement with previous work. However, we do not see any evidence for a dip in $v_{\rm c}$ at $R\approx 9$~kpc or $R\approx11$~kpc that was previously claimed by other studies \citep{Sofue2009, Kafle2012, Huang2016, McGaugh2018} and had been interpreted as potential signatures of spiral arms. We do see a mild but significant deviation from the straightly declining circular velocity curve at $R\approx 19-21$~kpc of $\Delta v\approx 15\,\rm km\,s^{-1}$. 
Differences in the estimation of $v_{\rm c}(R)$ between our work and previous studies could potentially be due to different tracer populations that have been used for the analyses. \citet{Kafle2012} for instance used blue horizontal branch stars in the halo, i.e. $|z|>4$~kpc, whereas our analysis is focused on stars within the Milky Way disk. 

Combined with other work for the mass distribution of the stellar components of the Milky Way, we estimated the NFW-profile of the dark matter halo, while keeping contributions from the baryonic components fixed. Our estimate of the virial mass, $M_{\rm vir}=(7.25\pm0.25)\cdot10^{11}\,M_{\odot}$, 
is significantly lower than what several previous studies suggest. The recently published analysis by \citet{Watkins2018} based on the kinematics of halo globular clusters determined by \gaia\ DR2 data, measures a virial mass of $M_{\rm vir} = 1.41^{+1.99}_{-0.52}\,M_{\odot}$ that is roughly twice as high than our value, although the measurements have large uncertainties \citep[see also][]{Piffl2014, Kuepper2015, BlandHawthornGerhard2016, Huang2016}. However, virial mass measurements that agree well with our estimate are presented by \citet{Bovy2012}, as well as \citet{Eadie2016} and \citet{Eadie2018}, whose analyses are similar to the one from \citet{Watkins2018} and also based on kinematic data of globular clusters. 

Our estimated local dark matter density is in good agreement within the uncertainties with the estimated values by \citet{Huang2016}, who measured $\rho_{\odot, \rm dm} =0.32\pm0.02 \,\rm GeV\,cm^{-3}$, and by \citet{Zhang2013}, who estimated $\rho_{\odot, \rm dm} =0.25\pm0.09 \,\rm GeV\,cm^{-3}$, although significantly lower than $\rho_{\odot, \rm dm} =0.542\pm0.042 \,\rm GeV\,cm^{-3}$ estimated by \citet{Bienayme2014}.
Additionally, our new measurement of the circular velocity curve covering large Galactocentric distances, enabled us to determine the Galactic radius beyond which the mass of the Milky Way is dominated by dark matter, which we estimate to be $R\gtrsim 14$~kpc. 
This value is in vague agreement with the result by \citet{BovyRix2013}, who found that the Milky Way's circular velocity is dominated by the stellar components at $R<10$~kpc. A similar study by \citet{Portail2017} obtains a distance of $R\approx 8$~kpc that marks the transition from the baryonic to dark matter dominated regime. 
Obviously, our inferred properties of the Milky Way's dark matter halo are dependent on the measurements of the baryonic mass components, which we keep fixed in our analysis. Indeed there is some disagreement in the literature about the exact stellar masses, and hence if the measurements from \citet{Pouliasis2017}, which we use in our analysis, are overestimated as suggested by \citet{Portail2017}, our results would be underestimating the influence of the dark matter halo.

\acknowledgments
It is a pleasure to thank Mike Fich, Ortwin Gerhard, Adrian Price-Whelan, Tobias Buck, Eddie Schlafly, and the participants in the MPIA Milky Way Group Meeting, as well as the \apogee\ telecon for helpful feedback and discussion. 

This project was developed in part at the 2018 \acronym{NYC} Gaia Sprint, hosted by the Center for Computational Astrophysics of the Flatiron Institute in New York City in June 2018.

This work has made use of data from the European Space Agency (\acronym{ESA}) mission \gaia\ (\url{https://www.cosmos.esa.int/gaia}), processed by the \gaia\ Data Processing and Analysis Consortium (\acronym{DPAC}, \url{https://www.cosmos.esa.int/web/gaia/dpac/consortium}). Funding for the \acronym{DPAC} has been provided by national institutions, in particular the institutions participating in the \gaia\ Multilateral Agreement. 

This publication makes use of data products from the Two Micron All Sky Survey, which is a joint project of the University of Massachusetts and the Infrared Processing and Analysis Center/California Institute of Technology, funded by the National Aeronautics and Space Administration and the National Science Foundation.

This publication makes use of data products from the Wide-field Infrared Survey Explorer, which is a joint project of the University of California, Los Angeles, and the Jet Propulsion Laboratory/California Institute of Technology, funded by the National Aeronautics and Space Administration.

The complete data and code of this work is available at \url{https://doi.org/10.5281/zenodo.1468053}
and \url{https://github.com/aceilers/spectroscopic_parallax}, respectively. 

\bibliography{literature_mw}

\end{document}